\documentclass[english,11pt]{article}
\usepackage[T1]{fontenc}
\usepackage[latin1]{inputenc}
\usepackage[pdftex]{graphicx}
\usepackage{verbatim}
\usepackage[width=\textwidth]{caption}
\usepackage{amsmath}
\usepackage{a4wide}
\usepackage{amssymb}
\usepackage[english]{babel}
\usepackage[]{float}
\usepackage[]{placeins}
\usepackage{flafter}
\usepackage[]{longtable}
\usepackage{booktabs}
\usepackage{verbatim}
\usepackage{amsopn}
\usepackage{dsfont}
\usepackage{color}
\usepackage{array}
\usepackage{natbib}
\usepackage{xr}
\usepackage{multirow}
\usepackage{lscape}
\usepackage{afterpage}
\usepackage{changepage}
\usepackage{enumerate}

\newtheorem{thm}{Theorem}[section]

\newtheorem{lem}{Lemma}[section]

\newtheorem{hyp}{Assumption}

\newcommand{\indep}{\perp \!\!\! \perp}
\newcommand{\eps}{\varepsilon}

\renewcommand{\citep}[1]{\citeauthor{#1}, \citeyear{#1}}
\newcolumntype{C}[1]{>{\centering\let\newline\\\arraybackslash\hspace{0pt}}m{#1}}

\date{\today}
\begin{document}

\title{Estimating the effect of treatments allocated by randomized waiting lists.\thanks{This paper originated from a comment an anonymous referee made to us when he/she reviewed \cite{behaghel2017}. We would like to thank him/her. We would also like to thank Hugo Botton for outstanding research assistance. We are very grateful to Chris Blattman and Jeannie Annan for making their data publicly available, and for patiently answering all our questions. We are also very grateful to Josh Angrist, Bart Cockx, Xavier D'Haultf\oe{}uille, Thomas Le Barbanchon, Chang Lee, Heather Royer, Doug Steigerwald, Chris Walters, members of the UCSB econometrics research group, and seminar participants at: the 2017 California Econometrics conference, the 2017 Labor and Education workshop of the NBER Summer Institute, Louvain-la-Neuve, the Paris School of Economics, UC Santa Barbara, UC San Diego, and Warwick for their helpful comments.}}
\author{ Cl\'{e}ment de Chaisemartin%
\thanks{University of California at Santa Barbara, clementdechaisemartin@ucsb.edu%
}
\and Luc Behaghel%
\thanks{Paris School of Economics, INRA, luc.behaghel@ens.fr%
}
}\maketitle  \vspace{-1cm}
\begin{abstract}
Oversubscribed treatments are often allocated using randomized waiting lists. Applicants are ranked randomly, and treatment offers are made following that ranking until all seats are filled. To estimate causal effects, researchers often compare applicants getting and not getting an offer. We show that those two groups are not statistically comparable. Therefore, the estimator arising from that comparison is inconsistent. We propose a new estimator, and show that it is consistent. Finally, we revisit an application, and we show that using our estimator can lead to sizably different results from those obtained using the commonly used estimator.
\end{abstract}



\newpage


\section{Introduction}\label{sec_intro}

Often times, some individuals who apply for a treatment are non-takers. They decline to get treated when they receive an offer, for instance because they then realize that their benefit from treatment is lower than they thought.
When a treatment is oversubscribed but some applicants are non-takers, an appealing way of allocating the available seats is to use randomized waitlists. First, applicants are ranked randomly. Then, if $S$ seats are available, an initial round of offers takes place, whereby the first $S$ applicants get an offer. If $r$ of them decline it, a subsequent round of offers takes place whereby the next $r$ applicants get an offer. Offers stop when all the seats have been filled. This allocation method is fair: each taker has the same probability of being treated; it is also efficient: no seat for treatment remains unused, despite the presence of non-takers. Therefore, oversubscribed treatments with non-takers are often allocated by randomized waitlists. We conducted a survey, and found 43 articles studying treatments allocated by randomized waitlists, ranging from charter schools in the USA to agricultural trainings
in Liberia. These treatments often have capacity constraints for various groups of applicants. For instance, a charter school may have 20 seats available in 7th grade and 25 seats in 8th grade. Then, a lottery takes place in each group.

\medskip
As applicants are ranked randomly, it may be possible to form two comparable groups with different likelihoods of getting an offer. One could then compare those two groups to estimate the effect of the treatment.
In practice, researchers have used two types of comparisons. Some researchers have compared applicants getting and not getting an initial offer, thus giving rise to the so-called initial-offer (IO) estimators. Other researchers have compared applicants ever and never getting an offer, thus giving rise to the so-called ever-offer (EO) estimators. When several lotteries were conducted, as in the charter school example above, researchers have often included waitlist fixed effects in their specifications, to ensure they compare applicants within and not across waitlists. In our survey, 22 articles used the EO estimator, 20 used the IO estimator, and a handful used other estimators. Overall, practices are not standardized.

\medskip
We start by showing that the expected proportion of takers is strictly greater among applicants ever getting an offer than among applicants never getting one. Intuitively, this is because
offers continue until sufficiently many takers have gotten an offer. 
Moreover, when waitlist fixed effects are included in the estimation, they induce an endogenous reweighting of waitlists that usually further increases this imbalance between the two groups,
as we explain in more detail in Section \ref{sec_example}. Then, we show that due to this imbalance, the EO estimator is inconsistent when the number of waitlists goes to infinity. In our survey, we find that articles using randomized waitlists often pool data from a large number of small waitlists, thus motivating the asymptotic sequence we consider. By contrast, if the number of applicants and takers per waitlist goes to infinity, the asymptotic bias of the EO estimator goes to 0. Accordingly, in simulations we find that the EO estimator is more biased when waitlists have fewer applicants and takers.

\medskip
It turns out that what creates the imbalance between applicants getting and not getting an offer is the fact that in each waitlist, the last applicant getting an offer must by construction be a taker. Indeed, we show that dropping that applicant in each waitlist is sufficient to restore the comparability between those two groups. 
Based on this result, we propose a new estimator of the treatment effect. It is built out of comparisons of applicants that get and do not get an offer in each waitlist, downweighting applicants that accept their offer by an amount equivalent to dropping one of them. Then, our estimator takes a weighted average of those within-waitlist comparisons, with a weighting scheme that avoids the endogeneous reweighting induced by the waitlist fixed effects. We refer to those estimators as the doubly-reweighted ever-offer estimators (DREO). We show that our estimator is consistent and asymptotically normal when the number of waitlists goes to infinity. 

\medskip
Contrary to subsequent-round offers, initial offers are only a function of applicants' random ranks in the waitlist. Therefore, applicants getting and not getting an initial offer are statistically comparable, and the IO estimator is also consistent. However, we find in simulations that the variance of that estimator is much larger than that of the DREO estimator, so using it will often result in large efficiency losses.

\medskip
We use our results to revisit \cite{blattman2016}, who studied the effects of an agricultural training. The DREO estimator is significantly and economically different from the EO estimator computed by the authors for some of the outcomes they considered.\footnote{A Stata adofile computing the DREO estimator is available from the authors' website.} 


\medskip
The remainder of the paper is organized as follows.
Section \ref{sec_example} uses a simple example to give the intuition of our results. Section \ref{sec_identification} presents our main results. Section \ref{sec_appli} presents our empirical application. Appendix \ref{sec_appendix_proofs} presents the proofs.
In our web appendix, we present our survey of articles that have used randomized waitlists, we show that some of the assumptions adopted in the paper can be relaxed, we present some simulations, and we revisit another application.

\section{Introducing the results through a simple example}\label{sec_example}

We start with a simple example.
We consider a waitlist where five applicants compete for three seats. Four applicants are takers ($T$) and one is a non-taker ($NT$), meaning that she will refuse to get treated if she gets an offer. Applicants are randomly ranked, and treatment offers are made following that ranking until all seats are filled. Table \ref{table_example1} displays the five possible orderings of the takers and the non-taker. For each ordering, applicants getting an offer are depicted in italics, while those not getting an offer are depicted in bold. In orderings 1 and 2, the first three applicants are takers, so offers stop after the third offer.
In orderings 3, 4, and 5, one of the first three applicants is a non-taker, so a fourth offer is made; then the next applicant is a taker so offers stop as the available seats have been filled.

\medskip
The first issue with the EO estimator is that, on average, applicants getting an offer bear a higher proportion of takers than applicants not getting an offer.
Each ordering has a 0.20 probability of being selected. Across the five orderings, the expected share of takers among applicants getting an offer is $0.2\times \left(1+1+3/4+3/4+3/4\right)=17/20.$ On the other hand, the expected share of takers among applicants not getting an offer is $0.2\times \left(1/2+1/2+1+1+1\right)=4/5.$
Intuitively, this imbalance arises because offers stop when sufficiently many takers have accepted an offer. This endogenous stopping rule creates a positive correlation between getting an offer and being a taker. When the average potential outcomes of takers and non takers differ,\footnote{This is often the case. \cite{abadie2002} and \cite{crepon2015} are just a few examples of the many papers that have found large differences between the average potential outcomes of takers and non-takers.} this imbalance implies that applicants getting and not getting an offer are not statistically comparable: those two groups have different average potential outcomes.
\begin{table}[H]
\begin{center}
\caption{Applicants \textit{getting} and \textbf{not getting} an offer in an example}
\label{table_example1}
\begin{tabular}{c |c |c |c | c}
Ordering 1 & Ordering 2 & Ordering 3 & Ordering 4 & Ordering 5\\
\textit{T} & \textit{T} & \textit{T} & \textit{T} & \textit{NT} \\
\textit{T} & \textit{T} & \textit{T} & \textit{NT} & \textit{T}\\
\textit{T} & \textit{T} & \textit{NT} & \textit{T} & \textit{T} \\
\textbf{T} & \textbf{NT} & \textit{T} & \textit{T} & \textit{T} \\
\textbf{NT} & \textbf{T} & \textbf{T} & \textbf{T} & \textbf{T}
\end{tabular}
\end{center}
\end{table}
The second issue with the EO estimator arises from the inclusion of fixed effects when pooling waitlists.
Assume that one pools waitlists that all have four takers, one non-taker, and three seats. In some waitlists, the realized ordering of takers and non-takers is ordering 1 in Table \ref{table_example1}, in other waitlists the realized ordering is ordering 2, etc. With several waitlists, it follows from, e.g., Equation (3.3.7) in \cite{angrist2008}, that the EO estimator with waitlist fixed effects is a weighted average of the
EO estimators in each waitlist,
that gives more weight to waitlists where the share of applicants getting an offer is closer to 1/2. In our example, 2/3 of applicants get an offer in waitlists with ordering 1 or 2, while 4/5 of applicants get an offer in waitlists with ordering 3, 4, or 5. Accordingly, waitlists with ordering 1 or 2 receive more weight. But those are precisely the waitlists where the proportion of takers among applicants getting an offer is the highest.
Therefore, the reweighting of waitlists induced by the fixed effects aggravates the over-representation of takers among applicants getting an offer.

\medskip
The DREO estimator we propose addresses those two issues. Firstly, in our example dropping the last taker getting an offer is sufficient to solve the endogenous stopping rule issue. Table \ref{table_example2} shows that then, the expected share of takers among applicants getting an offer is equal to $0.2\times \left(1+1+2/3+2/3+2/3\right)=4/5,$ the same as among applicants not getting an offer. Still, dropping the last taker getting an offer is arbitrary: dropping the first or the second would have the same effect.
Besides, doing so reduces the sample size and statistical precision. Instead,
one can give to the three of them a weight equal to 2/3: this reduces the expected share of takers among applicants getting an offer by the same amount as dropping one.
Secondly, instead of using fixed effects to pool waitlists, we simply take an average of the estimators in each waitlist, weighting waitlists proportionally to their number of applicants. These weights are independent of how many offers one has to make to fill the available seats, which solves the second issue of the EO estimator. Table \ref{table_example2} shows that this second reweighting is necessary. Even after downweighting takers getting an offer, including waitlist fixed effects would still lead to over-represent takers among applicants getting an offer. Indeed, doing so gives more weight to waitlists with ordering 1 or 2, where 1/2 of applicants get an offer, while those are the waitlists where the proportion of takers among applicants getting an offer is the highest.
\begin{table}[H]
\begin{center}
\caption{Applicants \textit{getting} and \textbf{not getting} an offer, dropping the last taker getting an offer}
\label{table_example2}
\begin{tabular}{c |c |c |c | c}
Ordering 1 & Ordering 2 & Ordering 3 & Ordering 4 & Ordering 5\\
\textit{T} & \textit{T} & \textit{T} & \textit{T} & \textit{NT} \\
\textit{T} & \textit{T} & \textit{T} & \textit{NT} & \textit{T}\\
 &  & \textit{NT} & \textit{T} & \textit{T} \\
\textbf{T} & \textbf{NT} &  &  &  \\
\textbf{NT} & \textbf{T} & \textbf{T} & \textbf{T} & \textbf{T}
\end{tabular}
\end{center}
\end{table}

\section{Main results}\label{sec_identification}

\subsection{Assumptions and parameter of interest}\label{subsec_assumptions}

Throughout the paper, we consider the following set-up.
\begin{hyp}\label{hyp:set-up}
(Set-up)
\begin{enumerate}[a)]
\item Applicants for a binary treatment are divided into $K$ mutually exclusive waitlists. For every $k\in \{1..K\}$, $N_k$ denotes the number of applicants in waitlist $k$. $N_k$ is non stochastic.
\item In each waitlist, $S_k$ seats are available, and are allocated as follows: applicants are ranked, and treatment offers are made following that order until $S_k$ applicants have accepted to get treated or all applicants have received an offer. $S_k$ is non stochastic.
\item Applicants that do not get an offer cannot get treated.
\end{enumerate}
\end{hyp}
In Section \ref{sec_extensions} in the Web appendix, we consider various extensions of this set-up. For instance, we show that our results remain unchanged if we allow for the possibility that some applicants manage to get treated even if they do not receive an offer. Similarly, we allow for the possibility that some applicants may participate in several waiting-lists, or that the treatment may not be binary.
But for now we focus on the basic set-up outlined in Assumption \ref{hyp:set-up}.

\medskip
Then, we assume that ranks are randomly assigned to applicants. Let $R_{ik}$ denote the rank assigned to applicant $i$ in waitlist $k$, let $L_k$ denote the number of applicants getting an offer in waitlist $k$, and let
$Z_{ik}=1\{R_{ik}\leq L_k\}$ denote whether applicant $i$ gets an offer, the so-called ever-offer instrument. Let $D_{ik}(1)$ denote her potential treatment if she gets an offer, and let
$D_{ik}$ denote her observed treatment. Under point c) of Assumption \ref{hyp:set-up}, $D_{ik}=Z_{ik}D_{ik}(1)$.
For every $d\in \{0,1\}$, let $Y_{ik}(d)$ denote her potential outcome if $D_{ik}=d$,\footnote{We implicitly assume that getting an offer does not have a direct effect on the outcome, the so-called exclusion restriction, see \cite{Angrist96}.} and let $Y_{ik}=Y_{ik}(D_{ik})$ denote her observed outcome.
Let $$\mathcal{P}_k=\left(\left(D_{1k}(1),Y_{1k}(0),Y_{1k}(1)\right),...,\left(D_{N_kk}(1),Y_{N_kk}(0),Y_{N_kk}(1)\right)\right)$$ be a vector stacking the potential treatments and outcomes of the applicants in waitlist $k$.
For any integer $j$, let $\Pi_j$ denote the set of permutations of $\{1..j\}$.
Let $\mathcal{R}_k=(R_{1k},...,R_{N_kk})$ denote the ranks assigned to applicants $1$ to $N_k$ in waitlist $k$.
\begin{hyp}\label{hyp:random_assignment_of_seats}
(Randomly assigned ranks)

\noindent
For all $k \in \{1..K\}$ and $(r_1,...,r_{N_k})\in \Pi_{N_k}$, $P(\mathcal{R}_k=(r_1,...,r_{N_k})|\mathcal{P}_k)=\frac{1}{N_k!}$.
\end{hyp}
Assumption \ref{hyp:random_assignment_of_seats} requires that the ranks assigned to applicants be independent of their potential treatments and outcomes, and uniformly distributed on $\Pi_{N_k}$. It implies that each applicant has the same probability of being in the first, second, ..., or last rank.

\medskip
Finally, we consider a last assumption.
Let applicants with $D_{ik}(1)=1$ (resp. $D_{ik}(1)=0$) be referred to as takers (resp. non-takers). For every $k \in \{1..K\}$, let $T_k=\sum_{i=1}^{N_k}D_{ik}(1)$ denote the number of takers in waitlist $k$.
\begin{hyp}\label{hyp:moretakersthanseats}
(Strictly more takers than seats)

\noindent
For every $k \in \{1..K\}$, $2\leq S_k<T_k$.
\end{hyp}
Assumption \ref{hyp:moretakersthanseats} requires that each waitlist have at least two seats. 
This can be assessed from the data, so waitlists with less than two seats can just be dropped.
Assumption  \ref{hyp:moretakersthanseats} also requires that each waitlist have strictly more takers than seats. This cannot be assessed from the data. When all the seats available in a waitlist get filled, it must be that $S_k\leq T_k$, but it is still possible that $S_k=T_k$: all applicants not getting an offer might be non-takers. Still, we show in Section \ref{subsec_testability} in the Web appendix that Assumption \ref{hyp:moretakersthanseats} is testable.

\medskip
Let $T=\sum_{k=1}^K T_k$ denote the total number of takers. Our parameter of interest is $$\Delta_K=E\left(\frac{1}{T}\sum_{(i,k):D_{ik}(1)=1}\left[Y_{ik}(1)-Y_{ik}(0)\right]\right),$$
the local average treatment effect of the takers.

\subsection{The Doubly Reweighted Ever Offer estimator}\label{subsec_DREO}

Let $N=\sum_{k=1}^K N_k$ and $\overline{N}=\frac{N}{K}$ respectively denote the total number of applicants and the average number of applicants per waitlist. Let $\mathcal{I}=\{(i,k)\in \mathbb{N}^2:i\in \{1..N_k\},k\in \{1..K\}\}$, and for every $(i,k)\in \mathcal{I}$, let $w_{ik}=1-\frac{Z_{ik}D_{ik}}{S_k}.$ $w_{ik}$ is equal to $1-\frac{1}{S_k}$ for applicants that get and accept an offer, and to $1$ for everyone else. As $S_k$ takers receive an offer in each waitlist, weighting applicants getting an offer by $w_{ik}$ decreases the share of takers among them by the same amount as dropping one taker, as illustrated in the numerical example in Section \ref{sec_example}.

\medskip
The DREO estimator of $\Delta_K$ is defined as
\begin{eqnarray}\label{eq:def_estimators}
\widehat{\Delta}=\frac{\frac{1}{K}\sum_{k=1}^K\frac{N_k}{\overline{N}}\left(\frac{1}{L_k-1}\sum_{i:Z_{ik}=1}w_{ik}Y_{ik}-\frac{1}{N_k-L_k}\sum_{i:Z_{ik}=0}Y_{ik}\right)}{\frac{1}{K}\sum_{k=1}^K\frac{N_k}{\overline{N}}\frac{1}{L_k-1}\sum_{i:Z_{ik}=1}w_{ik}D_{ik}}.
\end{eqnarray}
$\widehat{\Delta}$ can be computed through a 2SLS regression.
Let $L=\sum_{k=1}^KL_k$, and let
\begin{eqnarray*}
w_{ik}^{DR}&=&w_{ik}\left(Z_{ik}\times\frac{L-K}{N-K}\times\frac{N_k}{L_k-1}+\left(1-Z_{ik}\right)\times\frac{N-L}{N-K}\times\frac{N_k}{N_k-L_k}\right)
\end{eqnarray*}
be a weighting scheme combining $w_{ik}$ with propensity score reweighting.
One can show that $\widehat{\Delta}$ is equal to the coefficient of $D_{ik}$ in a 2SLS regression of $Y_{ik}$ on $D_{ik}$ using $Z_{ik}$ as the instrument, and weighted by $w_{ik}^{DR}$. Importantly, note that under Assumption \ref{hyp:set-up}, $S_k=\sum_{i=1}^{N_k}Z_{ik}D_{ik}$, so observing $(Z_{ik},D_{ik},Y_{ik})_{(i,k)\in \{1..N_k\}\times\{1..K\}}$ is sufficient to compute $\widehat{\Delta}$.

\medskip
Our main result relies on the following lemma:
\begin{lem}\label{lem:identification}
If Assumptions \ref{hyp:set-up}-\ref{hyp:moretakersthanseats} hold, then for all $k\in\{1..K\}$,
\begin{enumerate}[a)]
\item $E\left(\frac{1}{K}\sum_{k=1}^K\frac{N_k}{\overline{N}}\left(\frac{1}{L_k-1}\sum_{i:Z_{ik}=1}w_{ik}Y_{ik}-\frac{1}{N_k-L_k}\sum_{i:Z_{ik}=0}Y_{ik}\right)\right)=E\left(\frac{1}{N}\sum_{(i,k)\in \mathcal{I}}\left[Y_{ik}(D_{ik}(1))-Y_{ik}(0)\right]\right)$,
\item $E\left(\frac{1}{K}\sum_{k=1}^K\frac{N_k}{\overline{N}}\frac{1}{L_k-1}\sum_{i:Z_{ik}=1}w_{ik}D_{ik}\right)=E\left(\frac{1}{N}\sum_{(i,k)\in \mathcal{I}}D_{ik}(1)\right)$.
\end{enumerate}
\end{lem}
The intuition of point a) of the theorem goes as follows. As the numerical example in Section \ref{sec_example} illustrates, one can show that in each waitlist, $w_{ik}$-reweighted applicants getting an offer are statistically comparable to applicants not getting an offer.
Therefore, the only difference between these two groups is that one receives an offer and not the other one. Accordingly, $\frac{1}{L_k-1}\sum_{i:Z_{ik}=1}w_{ik}Y_{ik}-\frac{1}{N_k-L_k}\sum_{i:Z_{ik}=0}Y_{ik}$, the difference between the average outcome of the two groups, is an unbiased estimator of  $E\left(\frac{1}{N_k}\sum_{i=1}^{N_k}\left[Y_{ik}(D_{ik}(1))-Y_{ik}(0)\right]\right),$ the intention to treat effect of getting an offer on applicants' outcome in waitlist $k$. The numerator of $\widehat{\Delta}$ is an average of those unbiased within-waitlist comparisons, that gives to each waitlist a weight proportional to its number of applicants. Therefore, this numerator is an unbiased estimator of $E\left(\frac{1}{N}\sum_{(i,k)\in \mathcal{I}}\left[Y_{ik}(D_{ik}(1))-Y_{ik}(0)\right]\right)$, the intention to treat effect
among all applicants. The intuition of point b) is similar. 

\medskip
We now derive the asymptotic distribution of $\widehat{\Delta}$. In our survey of articles that have used randomized waitlists, the median number of waitlists used in the analysis is equal to 64. Therefore, we consider a sequence where $K$, the number of waitlists, goes to infinity.
An alternative would be to consider a sequence where the number of applicants per waitlist goes to infinity, but in our survey the median of waitlists divided by applicants per waitlist is equal to 1.9, so the former asymptotic may be more appropriate in a majority of applications.
For all $k\in\{1..K\}$, let $RF_k=\frac{N_k}{\overline{N}}\left[\frac{1}{L_k-1}\sum_{i:Z_{ik}=1}w_{ik}Y_{ik}-\frac{1}{N_k-L_k}\sum_{i:Z_{ik}=0}Y_{ik}\right]$ and
$FS_k=\frac{N_k}{\overline{N}}\frac{1}{L_k-1}\sum_{i:Z_{ik}=1}w_{ik}D_{ik}.$
Let also $FS=\underset{K\rightarrow +\infty}{\lim}\frac{1}{K}\sum_{k=1}^KE\left(FS_k\right)$ and $\Delta=\underset{K\rightarrow +\infty}{\lim}\Delta_K$, where Assumption \ref{hyp:iid} below ensures that those limits exist.
Finally, for all $k$ let  $\Lambda_k=\frac{RF_k-\Delta FS_k}{FS}$.
\begin{hyp}\label{hyp:iid}
(Technical assumptions to derive the asymptotic distribution of $\widehat{\Delta}$)
\begin{enumerate}[a)]
\item The vectors $(\mathcal{P}_k,\mathcal{R}_k)_{1\leq k\leq K}$ are mutually independent.
\item $T \indep \frac{1}{T}\sum_{(i,k):D_{ik}(1)=1}\left[Y_{ik}(1)-Y_{ik}(0)\right]$.
\item For every $k$, $N_k\leq N^+$, for some integer $N^+$.
\item For every $k$, $E\left(RF_k^{4}\right)<+\infty$. 
\item $\frac{1}{K}\sum_{k=1}^KE\left(RF_k\right)$, $\frac{1}{K}\sum_{k=1}^KE\left(FS_k\right)$, $\Delta_K$,
    $\frac{1}{K}\sum_{k=1}^KE\left(RF^2_k\right)$, $\frac{1}{K}\sum_{k=1}^KE\left(FS^2_k\right)$,
    $\frac{1}{K}\sum_{k=1}^KV\left(RF_k\right)$, $\frac{1}{K}\sum_{k=1}^KV\left(FS_k\right)$,
    $\frac{1}{K}\sum_{k=1}^KE\left(RF_kFS_k\right)$,
    $\frac{1}{K}\sum_{k=1}^KE\left(\left(RF_k-E\left(RF_k\right)\right)^{4}\right)$, $\frac{1}{K}\sum_{k=1}^KE\left(\left(FS_k-E\left(FS_k\right)\right)^{4}\right)$,\\ and $\frac{1}{K}\sum_{k=1}^KE\left(\left(\Lambda_k-E\left(\Lambda_k\right)\right)^{4}\right)$   converge towards finite limits when $K\rightarrow +\infty$.
\item $\sum_{k=1}^{+\infty}\frac{V(RF_k)}{k^2}<+\infty$ and $\sum_{k=1}^{+\infty}\frac{V(RF_k^2)}{k^2}<+\infty$.
\end{enumerate}
\end{hyp}
Typically, the lotteries determining applicants' ranks are independent across waitlists, so by design the vectors $(\mathcal{R}_k)_{1\leq k\leq K}$ are mutually independent, and $(\mathcal{R}_k)_{1\leq k\leq K}$ is independent of $(\mathcal{P}_k)_{1\leq k\leq K}$. Then, point a) of Assumption \ref{hyp:iid} only requires that the vectors $(\mathcal{P}_k)_{1\leq k\leq K}$ be mutually independent. This is often plausible, for instance when the waitlists correspond to different schools. If point a) is not plausible, then Theorem \ref{thm:inference} below still holds conditional on applicants' potential treatments and outcomes, as in \cite{abadie2014} or \cite{li2017}. Point b) requires that the number of takers be independent of their average treatment effect.
If point b) does not hold, Theorem \ref{thm:inference} below still holds, except that $\Delta_K$ has to be replaced by $\Delta$, and more technical assumptions have to be made. Point c) requires that the number of applicants per waitlist be uniformly bounded by some constant $N^+$. Points d), e), and f) are technical conditions ensuring we can apply Liapunov's central limit theorem and Kolmogorov's strong law to $(RF_k)_{k\in
\mathbb{N}}$, $(FS_k)_{k\in
\mathbb{N}}$, and $(\Lambda_k)_{k\in
\mathbb{N}}$. One can show that d) and f) hold if the potential outcomes $Y_{ik}(0)$ and $Y_{ik}(1)$ have a bounded support.

\medskip
Let $\sigma^2=\underset{K\rightarrow +\infty}{\lim}\frac{1}{K}\sum_{k=1}^K V(\Lambda_k),$ $\sigma^2_+=\underset{K\rightarrow +\infty}{\lim}\left[\frac{1}{K}\sum_{k=1}^K E(\Lambda_k^2)-\left(\frac{1}{K}\sum_{k=1}^KE\left(\Lambda_k\right)\right)^2\right],$ \\
$\widehat{\Lambda}_k=\frac{RF_k-\widehat{\Delta}FS_k}{\frac{1}{K}\sum_{k=1}^KFS_k}$, and $\widehat{\sigma}^2_+=\frac{1}{K}\sum_{k=1}^K \left(\widehat{\Lambda}_k-\frac{1}{K}\sum_{j=1}^K\widehat{\Lambda}_j\right)^2.$
\begin{thm}\label{thm:inference}
If Assumptions \ref{hyp:set-up}-\ref{hyp:iid} hold, $\sqrt{K}\left(\widehat{\Delta}-\Delta_K\right)~{\overset{d}{\longrightarrow}}~ \mathcal{N}\left(0,\sigma^2\right)$ and $\widehat{\sigma}^2_+~{\overset{p}{\longrightarrow}}~\sigma^2_+\geq \sigma^2.$
\end{thm}
Theorem \ref{thm:inference} implies that $\widehat{\Delta}$ is an asymptotically normal estimator of $\Delta_K$ when the number of waitlists goes to infinity. As is usually the case for estimators constructed using independent but not identically distributed random variables (see e.g. \citep{liu1995}), the asymptotic variance $\sigma^2$ of $\widehat{\Delta}$ can only be conservatively estimated: we provide a consistent estimator of $\sigma^2_+$, an upper bound of $\sigma^2$. That estimator can then be used to build conservative confidence intervals for $\Delta_K$.\footnote{Conservative variance estimators also arise in other articles studying treatment effect estimation in randomized experiments (see e.g. \citep{Neyman1923}).} When all the $\Lambda_k$ have the same expectation, something that for instance happens when all waitlists have the same number of applicants, the same expectation of the proportion of takers, and the same expectations of takers' and non takers' potential outcomes, $\sigma^2_+=\sigma^2$ so those confidence intervals are exact. Finally, in simulations shown in Section \ref{subsec_simu_inference} of the Web appendix, we find that the asymptotic distribution in Theorem \ref{thm:inference} approximates the distribution of $\widehat{\Delta}$ well if 20 waitlists or more are used in the analysis. This suggests that articles using more than 20 waitlists may rely on Theorem \ref{thm:inference} for inference, while articles using less than 20 waitlists may not.

\subsection{Comparison with the Ever Offer and Initial Offer estimators}\label{subsec_EOIO}

\subsubsection{Comparison with the Ever Offer estimator}

Let $\widehat{\beta}^{E}_{FE}$ be the coefficient of $D_{ik}$ in a 2SLS regression of $Y_{ik}$ on $D_{ik}$ and waitlist fixed effects, using $Z_{ik}$ as the instrument for $D_{ik}$.
We refer to $\widehat{\beta}^{E}_{FE}$ as the EO estimator.
The derivation of its limit relies on Assumption \ref{hyp:plim_EO}, another technical assumption, that is stated in the proofs. Assumption \ref{hyp:plim_EO} is similar to points d) to f) in Assumption \ref{hyp:iid}, and it ensures that the limits in the definition of $w_k$ and $B$ below exist.
Let
\begin{align*}
w_k=&\frac{\frac{S_k\left(N_k-S_k\frac{N_k+1}{T_k+1}\right)}{N_k}}{\underset{K\rightarrow +\infty}{\lim}\frac{1}{K}\sum_{j=1}^K
E\left(\frac{S_j\left(N_j-S_j\frac{N_j+1}{T_j+1}\right)}{N_j}\right)},\\
B=&\frac{\underset{K\rightarrow +\infty}{\lim}\frac{1}{K}\sum_{k=1}^KE\left(\frac{S_k\left(N_k-T_k\frac{N_k+1}{T_k+1}\right)}{N_k}\left[\frac{1}{T_k}\sum_{i:D_{ik}(1)=1}Y_{ik}(0)-\frac{1}{N_k-T_k}\sum_{i:D_{ik}(1)=0}Y_{ik}(0)\right]\right)}{\underset{K\rightarrow +\infty}{\lim}\frac{1}{K}\sum_{k=1}^KE\left(\frac{S_k\left(N_k-S_k\frac{N_k+1}{T_k+1}\right)}{N_k}\right)}.
\end{align*}
\begin{thm}\label{thm:inconsistencyEO}
If Assumptions \ref{hyp:set-up}-\ref{hyp:plim_EO} hold,
\begin{equation}\label{eq:plimEO}
\widehat{\beta}^{E}_{FE}~{\overset{p}{\longrightarrow}}~\underset{K\rightarrow +\infty}{\lim}\frac{1}{K}\sum_{k=1}^K E\left(w_k\frac{1}{T_k}\sum_{i:D_{ik}(1)=1}\left[Y_{ik}(1)-Y_{ik}(0)\right]\right)+B.
\end{equation}
\end{thm}
Under Assumptions \ref{hyp:set-up}-\ref{hyp:plim_EO}, $\widehat{\beta}^{E}_{FE}$ converges towards the sum of two terms. The first is a weighted average of the LATEs of takers in each waitlist. If those LATEs vary across waitlists, this weighted average is not equal to the LATE of all takers, because it overrepresents waitlists with a ratio of seats to takers closer to 1/2.\footnote{This can be seen from the fact that  $\frac{S_k\left(N_k-S_k\frac{N_k+1}{T_k+1}\right)}{N_k}=T_k\frac{S_k}{T_k}\left(1-\frac{S_k}{(T_k+1)N_k/(N_k+1)}\right)\approx T_k\frac{S_k}{T_k}\left(1-\frac{S_k}{T_k}\right)$.} The second term, $B$, is a bias term. As explained in Section \ref{sec_example}, this bias arises from the endogenous stopping of offers in each waitlist, and from the waitlist fixed effects.

\medskip
We start by performing comparative statics on $\left|B\right|$, assuming that waitlists are homogeneous: there exist real numbers $N_0$, $T_0$, $S_0$, and $\Delta_{Y(0)}$ such that for all $k$, $N_k=N_0$, $T_k=T_0$, $S_k=S_0$, and $E\left(\left[\frac{1}{T_k}\sum_{i:D_{ik}(1)=1}Y_{ik}(0)-\frac{1}{N_k-T_k}\sum_{i:D_{ik}(1)=0}Y_{ik}(0)\right]\right)=\Delta_{Y(0)}$. Then,
\begin{equation}\label{eq:asymptotic_bias_EO_identical_waitlists}
B=
\frac{1-\left(1+\frac{1}{N_0}\right)\frac{t_0}{t_0+\frac{1}{N_0}}}{1-\left(1+\frac{1}{N_0}\right)\frac{s_0}{t_0+\frac{1}{N_0}}}\Delta_{Y(0)},
\end{equation}
where $t_0=T_0/N_0$ and $s_0=S_0/N_0$ respectively denote the proportion of takers and the ratio of seats to applicants in the waitlist. One can show that the right hand side of \eqref{eq:asymptotic_bias_EO_identical_waitlists} is decreasing in $N_0$, decreasing in $t_0$, increasing in $s_0$, and increasing in $\left|\Delta_{Y(0)}\right|$.

\medskip
Then, we study how waitlists' heterogeneity affects $\left|B\right|$. Let $(S^a_0,S^b_0)\in\{2..T_0-1\}^2$, let $(T^a_0,T^b_0)\in\{3..N_0\}^2$, and let $\Delta_{Y(0),k}=E\left[\frac{1}{T_k}\sum_{i:D_{ik}(1)=1}Y_{ik}(0)-\frac{1}{N_k-T_k}\sum_{i:D_{ik}(1)=0}Y_{ik}(0)\right]$. The three following results hold:
\begin{enumerate}
\item If $\left(N_k,T_k,\Delta_{Y(0),k}\right)=\left(N_0,T_0,\Delta_{Y(0)}\right)$ for all $k$, $\left|B\right|$ is larger if $\alpha$\% of the waitlists have $S^a_0$ seats  and $(1-\alpha)$\% have $S^b_0$ seats than if all of them have $\alpha S^a_0+(1-\alpha)S^b_0$ seats.
\item If $\left(N_k,S_k,\Delta_{Y(0),k}\right)=\left(N_0,S_0,\Delta_{Y(0)}\right)$ for all $k$, $\left|B\right|$ is larger if $\alpha$\% of the waitlists have $T^a_0$ takers and $(1-\alpha)$\% have $T^b_0$ takers than if all of them have $\alpha T^a_0+(1-\alpha)T^b_0$ takers.
\item If $\left(\frac{T_k}{N_k},\frac{S_k}{N_k},\Delta_{Y(0),k}\right)=\left(t_0,s_0,\Delta_{Y(0)}\right)$ for all $k$, $\left|B\right|$ is larger if $\alpha$\% of the waitlists have $N^a_0$ applicants and $(1-\alpha)$\% have $N^b_0$ applicants than if all have $\alpha N^a_0+(1-\alpha)N^b_0$ applicants.
\end{enumerate}
Overall, $\left|B\right|$ seems to be higher when waitlists have heterogeneous numbers of applicants, takers, and seats. The impact of waitlists' heterogeneity on $\left|B\right|$ can be large. For instance, if $\left(N_k,S_k,\Delta_{Y(0),k}\right)=\left(40,20,\Delta_{Y(0)}\right)$, $\left|B\right|$ is 17.1\% larger if $50$\% of waitlists have $25$ takers and $50$\% have $35$ takers than if all have $30$ takers.

\subsubsection{Comparison with the Initial Offer estimator}

Let $Z'_{ik}=1\{R_{ik}\leq S_k\}$ be an indicator for applicants in the initial round of offers, the so-called initial-offer instrument. Let $S=\sum_{k=1}^{K}S_k$. Let 
$w^{I}_{ik}=Z'_{ik}\times\frac{S}{N}\times\frac{N_k}{S_k}+\left(1-Z'_{ik}\right)\times\frac{N-S}{N}\times\frac{N_k}{N_k-S_k}$
be the propensity score weights attached to initial offers.
Let $\widehat{\beta}^{I}_{PS}$ be the coefficient of $D_{ik}$ in a 2SLS regression of $Y_{ik}$ on $D_{ik}$, using $Z'_{ik}$ as the instrument, and weighted by $w^{I}_{ik}$. We call $\widehat{\beta}^{I}_{PS}$ the IO estimator.

\medskip
Under Assumptions \ref{hyp:set-up}-\ref{hyp:random_assignment_of_seats} and a technical condition similar to Assumption \ref{hyp:iid}, $\sqrt{K}\left(\widehat{\beta}^{I}_{PS}-\Delta_K\right)$ converges towards a normal distribution. Contrary to $Z_{ik}$, $Z'_{ik}$ is only a function of applicants' random numbers and of the number of seats in their waitlist. Thus, it satisfies the random instrument assumption in \cite{Imbens1994}. Under Assumption \ref{hyp:set-up}, it also satisfies the monotonicity condition therein. Then, one can show that $\widehat{\beta}^{I}_{PS}$ is an asymptotically normal estimator of the LATE of applicants complying with an initial offer. As those are a random subset of the takers, this LATE is equal to $\Delta_K$.

\medskip
However, using $\widehat{\beta}^{I}_{PS}$ instead of $\widehat{\Delta}$ may result in a large loss of precision. In simulations shown in Section \ref{subsec_simu_baseline} in our Web appendix, we find that the variance of $\widehat{\Delta}$ is between 27.6 and 57.3\% smaller than that of $\widehat{\beta}^{I}_{PS}$, depending on the design we consider. This may reflect the fact that $\widehat{\beta}^{I}_{PS}$'s first stage is lower than that of $\widehat{\Delta}$, as some takers that do not get an initial offer get one in a subsequent round and get treated.

\section{Application to \cite{blattman2016}}\label{sec_appli}

After the second Liberian civil war, some ex-fighters started engaging in illegal activities, and working abroad as mercenaries.
\cite{blattman2016}\footnote{\cite{blattman2016} is one of the few articles in our survey in Section \ref{sec_litreview} whose data is not proprietary.} study the effect of an agricultural training on their employment and on their social networks. By improving their labor market opportunities, the program hoped to reduce their interest in illegal and mercenary activities, and to sever their relationships with other ex-combatants. To allocate the treatment, the authors divided applicants into 70 waitlists, according to the training site they applied for, their former military rank, and their community. In each waitlist, they randomly ranked applicants, and offers were made following that ranking until the seats available were filled.

\medskip
\cite{blattman2016} estimate the training's effect on 62 outcomes, that are either applicants' answers to survey questions, or indexes averaging their answers to several related questions. To preserve space, we only consider some outcomes. Here are the rules we used to make our selection: we chose indexes rather than questions averaged into an index; among questions not averaged into an index, we discarded those asking applicants to give a subjective opinion;
finally, we discarded a few measures the authors did not comment on in the paper. 
We end up with four measures of employment, 
one measure of applicants' interest in working as mercenaries, and
five measures of their social network. 

\medskip
For each outcome, Table \ref{table_blattman} below shows the EO estimator computed by the authors, and the DREO estimator computed with the same controls as those used by the authors.\footnote{The DREO estimator with controls is defined in Section \ref{subsec_covariates} of the Web Appendix.}  An estimate of $\widehat{\sigma}_+/\sqrt{K-1}$ is shown next to each DREO estimator.\footnote{To account for the controls included in the estimation, $Y_{ik}$ and $D_{ik}$ are regressed on the controls, and then the residuals from those two regressions are used instead of $Y_{ik}$ and $D_{ik}$ in the computation of $\widehat{\sigma}_+$.} The standard errors next to each EO estimator are computed using a bootstrap clustered at the waitlist level.\footnote{It follows from Theorem 2 in \cite{liu1995} that under point a) of Assumption \ref{hyp:iid} and the technical conditions therein, this bootstrap yields a conservative estimate of the variance of the EO estimator.} The table also shows the p-value of a t-test that the EO and DREO estimators are equal, also computed using the bootstrap. Finally, the table shows the estimated difference between the mean of $Y_{ik}(0)$ among non-takers and takers. The EO and DREO estimators are close for all employment outcomes, but they significantly differ for three of the other outcomes. For those outcomes, the differences between the estimators are large: for applicants' interest in mercenary work, the DREO estimator is 51.0\% larger in absolute value than the EO one; for applicants' relations with their ex-commanders, the DREO estimator is 47.4\% larger, and it is statistically significant while the EO estimator is not; for applicants' social network quality, the DREO estimator is three times larger, but none of the two estimators is significant. For the first two outcomes, the estimated difference between the mean of $Y_{ik}(0)$ of takers and non-takers is large, which may explain why the EO and DREO estimators differ.
\begin{table}[H]
\begin{adjustwidth}{0in}{0in}
\begin{center}
\caption{Estimators of the LATE in \cite{blattman2016}}
\label{table_blattman}

\begin{tabular}{lcccc} \hline \hline
 & EO (s.e.) & DREO (s.e.) & EO=DREO & $\Delta_{Y(0)}$ (s.e.) \\  \hline
Works in agriculture &        0.155 (0.041) &  0.167 (0.037) &           0.294 &                       0.020 (0.045) \\
Hours illegal work &       -3.697 (1.783)  &       -3.188 (1.614) &          0.264 &                       -2.807 (3.126) \\
Hours farming work &        4.090 (1.473) &        4.319 (1.472) &             0.654 &                       3.070 (2.219) \\
Income index &        0.157 (0.081) &        0.169 (0.069) &                0.663 &                     -0.087  (0.140) \\
Interest mercenary work &       -0.239 (0.136) &       -0.361 (0.155) &          0.041 &                        0.307 (0.226) \\
Relations ex-combatants &        0.073 (0.085) &        0.050 (0.097) &                0.501 &                       -0.079 (0.149) \\
Relations ex-commanders &       -0.154 (0.114) &       -0.227 (0.109) &         0.026 &              0.251 (0.141) \\
Social network quality &        0.027  (0.074) &        0.082 (0.068) &               0.092 &               -0.041 (0.128) \\
Social support &        0.188 (0.087) &        0.161 (0.089) &                0.345 &                       -0.165  (0.135)\\
Relationships families &        0.133 (0.079) &        0.161 (0.079) &            0.228 &               -0.059  (0.143)\\
N &     1,025 &     1,016 &             &        \\
\hline \hline \end{tabular}

\end{center}
\end{adjustwidth}
\footnotesize{\textit{Notes}. Columns 2 and 3 show the EO and DREO estimators in \cite{blattman2016}, for the outcome variables in Column 1,
and with the same controls as in \cite{blattman2016}. The EO estimators are computed using all the waitlists, while the DREO estimators are computed excluding one waitlist that had less than two seats. An estimate of $\widehat{\sigma}_+/\sqrt{K-1}$ accounting for the controls included in the estimation is shown next to each DREO estimator, between parentheses. Standard errors computed using a bootstrap clustered at the waitlist level are shown next to each EO estimator. Column 4 shows the p-value of a t-test that the EO and DREO estimators are equal, using the bootstrap to compute the standard error of the difference between the two estimators. Column 5 shows the estimated difference between the mean of $Y_{ik}(0)$ among takers and non-takers, as well as the bootstrap standard error of that difference.}
\end{table}

\section{Conclusion}\label{sec_conclu}

When the seats available for a treatment are allocated using randomized waitlists, we show that applicants getting and not getting an offer are not statistically comparable. Accordingly, a commonly used estimator of the treatment effect, the ever-offer estimator, is inconsistent when the number of waitlists goes to infinity. We propose a new estimator, the doubly-reweighted ever-offer (DREO) estimator, and we show that it is consistent and asymptotically normal. Simulations show that the DREO estimator is more efficient than another consistent estimator, the initial-offer estimator. Overall, we recommend that practitioners use the DREO estimator when they analyze randomized waitlists.

\newpage
\bibliography{biblio}

\newpage
\appendix

\section{Proofs}\label{sec_appendix_proofs}

The next lemma shows that the expectation of the average of any function of potential treatments and outcomes is the same among $w_{ik}$-reweighted applicants getting an offer and those not getting an offer. $\forall (i,k)\in \mathcal{I}$, let $P_{ik}=(D_{ik}(1),Y_{ik}(0),Y_{ik}(1))$.
\begin{lem}\label{lem:balancing}
If Assumptions \ref{hyp:set-up}-\ref{hyp:moretakersthanseats} hold, then  $\forall k\in \{1..K\}$ and for any function $\phi: \mathbb{R}^3\mapsto \mathbb{R}$,
\begin{eqnarray*}
E\left(\frac{1}{L_k-1}\sum_{i:Z_{ik}=1}w_{ik}\phi\left(P_{ik}\right)\middle|~\mathcal{P}_k\right)&=&E\left(\frac{1}{N_k-L_k}\sum_{i:Z_{ik}=0}\phi\left(P_{ik}\right)\middle|~\mathcal{P}_k\right)=\frac{1}{N_k}\sum_{i=1}^{N_k}\phi\left(P_{ik}\right).
\end{eqnarray*}
\end{lem}

\subsection*{Proof of Lemma \ref{lem:balancing}}

We start by showing that
\begin{equation}\label{eq:lem_sup0}
E\left(\frac{1}{L_k-1}\sum_{i:Z_{ik}=1}w_{ik}\phi\left(P_{ik}\right)\middle|~\mathcal{P}_k\right)=\frac{1}{N_k}\sum_{i=1}^{N_k}\phi\left(P_{ik}\right).
\end{equation}
First, we show that  \eqref{eq:lem_sup0} holds when $\mathcal{P}_k$ is such that $T_k<N_k$. Then, we have
\begin{eqnarray}
&&E\left(\frac{1}{L_k-1}\sum_{i:Z_{ik}=1}w_{ik}\phi\left(P_{ik}\right)\middle|~\mathcal{P}_k\right)\nonumber\\
&=&E\left(\sum_{i=1}^{N_k}\frac{1}{L_k-1}\left(1-\frac{D_{ik}(1)}{S_k}\right)\phi(P_{ik})1\{R_{ik}\leq L_k\}\middle|~\mathcal{P}_k\right)\nonumber\\
&=&\sum_{i=1}^{N_k}\left(1-\frac{D_{ik}(1)}{S_k}\right)\phi(P_{ik})E\left(\frac{1}{L_k-1}1\{R_{ik}\leq L_k\}\middle|~\mathcal{P}_k\right)\nonumber\\
&=&\sum_{i=1}^{N_k}\left(1-\frac{D_{ik}(1)}{S_k}\right)\phi(P_{ik})\sum\limits_{l=S_k}^{N_k-T_k+S_k}P(L_k=l|\mathcal{P}_k)\frac{1}{l-1}E\left(1\{R_{ik}\leq l\}\middle|~L_k=l,\mathcal{P}_k\right)\nonumber\\
&=&\sum_{i=1}^{N_k}\left(1-\frac{D_{ik}(1)}{S_k}\right)\phi(P_{ik})\sum\limits_{l=S_k}^{N_k-T_k+S_k}\frac{\binom{l-1}{S_k-1}\binom{N_k-l}{T_k-S_k}}{\binom{N_k}{T_k}}\frac{1}{l-1}E\left(1\{R_{ik}\leq l\}\middle|~L_k=l,\mathcal{P}_k\right)\nonumber\\
&=&\sum_{i=1}^{N_k}\left(1-\frac{D_{ik}(1)}{S_k}\right)\phi(P_{ik})\sum\limits_{l=S_k}^{N_k-T_k+S_k}\frac{\binom{l-1}{S_k-1}\binom{N_k-l}{T_k-S_k}}{\binom{N_k}{T_k}}\frac{1}{l-1}\left(D_{ik}(1)\frac{S_k}{T_k}+(1-D_{ik}(1))\frac{l-S_k}{N_k-T_k}\right)\nonumber\\
&=&\frac{1}{N_k}\sum_{i=1}^{N_k}\phi(P_{ik})\left(D_{ik}(1)\sum\limits_{l=S_k}^{N_k-T_k+S_k}\frac{\binom{l-1}{S_k-1}\binom{N_k-l}{T_k-S_k}\frac{S_k-1}{l-1}}{\binom{N_k}{T_k}\frac{T_k}{N_k}}+(1-D_{ik}(1))\sum\limits_{l=S_k+1}^{N_k-T_k+S_k}\frac{\binom{l-1}{S_k-1}\binom{N_k-l}{T_k-S_k}\frac{l-S_k}{l-1}}{\binom{N_k}{T_k}\frac{N_k-T_k}{N_k}}\right)\nonumber\\
&=&\frac{1}{N_k}\sum_{i=1}^{N_k}\phi(P_{ik})\left(D_{ik}(1)\sum\limits_{l=S_k-1}^{N_k-T_k+S_k-1}\frac{\binom{l-1}{S_k-2}\binom{N_k-1-l}{T_k-S_k}}{\binom{N_k-1}{T_k-1}}+(1-D_{ik}(1))\sum\limits_{l=S_k}^{N_k-1-T_k+S_k}\frac{\binom{l-1}{S_k-1}\binom{N_k-1-l}{T_k-S_k}}{\binom{N_k-1}{T_k}}\right)\nonumber\\
&=&\frac{1}{N_k}\sum_{i=1}^{N_k}\phi(P_{ik})\label{eq:lem1}.
\end{eqnarray}
The first equality follows from the definitions of $w_{ik}$, $Z_{ik}$, and $D_{ik}$. The second equality holds because $D_{ik}(1)$ and $\phi(P_{ik})$ are functions of $\mathcal{P}_k$, $N_k$ and $S_k$ are non stochastic, and the conditional expectation is linear. The third follows from the law of iterated expectations, and the fact that $L_k$ is included between $S_k$ and $N_k-T_k+S_k$ under Assumptions \ref{hyp:set-up} and \ref{hyp:moretakersthanseats}.

\medskip
Then, under Assumption \ref{hyp:set-up}, having $L_k=l$ is equivalent to having $S_k-1$ takers with $R_{ik}\leq l-1$, one with $R_{ik}=l$, and $T_k-S_k$ with $R_{ik}\geq l+1$. $\binom{l-1}{S_k-1}\binom{N_k-l}{T_k-S_k}T_k!(N_k-T_k)!$ possible values of $\mathcal{R}_k$ satisfy these constraints. Under Assumption \ref{hyp:random_assignment_of_seats}, conditional on $\mathcal{P}_k$ each of those values has a probability $\frac{1}{N_k!}$ of being realized. Hence the fourth equality.

\medskip
Then,
\small
\begin{eqnarray}\label{eq:lem_sup1}
E\left(1\{R_{ik}\leq l\}\middle|~L_k=l,\mathcal{P}_k\right)&=&D_{ik}(1)E\left(1\{R_{ik}\leq l\}\middle|~L_k=l,D_{ik}(1)=1,\mathcal{P}_k\setminus D_{ik}(1) \right)\nonumber\\
&+&(1-D_{ik}(1))E\left(1\{R_{ik}\leq l\}\middle|~L_k=l,D_{ik}(1)=0,\mathcal{P}_k\setminus D_{ik}(1)\right).
\end{eqnarray}
\normalsize
Conditional on $L_k=l$, $S_k$ takers out of $T_k$ satisfy $R_{ik}\leq l$, and Assumption \ref{hyp:random_assignment_of_seats} ensures that each taker has the same probability of satisfying this condition, so
\begin{equation}\label{eq:lem_sup2}
E\left(1\{R_{ik}\leq l\}\middle|~L_k=l,D_{ik}(1)=1,\mathcal{P}_k\setminus D_{ik}(1)\right)=\frac{S_k}{T_k}.
\end{equation}
Similarly, conditional on $L_k=l$ and $T_k<N_k$, $l-S_k$ non-takers out of $N_k-T_k$ satisfy $R_{ik}\leq l$, and Assumption \ref{hyp:random_assignment_of_seats} ensures that each has the same probability of satisfying this condition, so
\begin{equation}\label{eq:lem_sup3}
E\left(1\{R_{ik}\leq l\}\middle|~L_k=l,D_{ik}(1)=0,\mathcal{P}_k\setminus D_{ik}(1)\right)=\frac{l-S_k}{N_k-T_k}.
\end{equation}
Plugging  \eqref{eq:lem_sup2} and \eqref{eq:lem_sup3} into \eqref{eq:lem_sup1} yields the fifth equality. The sixth and seventh equalities follow after some algebra. 

\medskip
Then, we prove the eighth equality. Before that, note that $T_k<N_k$ and Assumption \ref{hyp:moretakersthanseats} ensure that $1\leq S_k-1 \leq T_k-1\leq N_k-1$ and $1\leq S_k \leq T_k\leq N_k-1$, thus ensuring that all the quantities that follow are well-defined.
There are $\binom{N_k-1}{T_k-1}$ ways of distributing $T_k-1$ units over $N_k-1$ ranks. The rank of the $S_k-1$th unit must be included between $S_k-1$ and $N_k-T_k+S_k-1$, and for every $l\in \{S_k-1..N_k-T_k+S_k-1\}$, there are $\binom{l-1}{S_k-2}\binom{N_k-1-l}{T_k-S_k}$ ways of distributing those $T_k-1$ units while having that the $S_k-1$th unit is at the
 $l$th rank. Therefore,
\begin{equation}\label{eq:lem_sup4}
\sum\limits_{l=S_k-1}^{N_k-T_k+S_k-1}\binom{l-1}{S_k-2}\binom{N_k-1-l}{T_k-S_k}=\binom{N_k-1}{T_k-1}.
\end{equation}
Similarly, 
when distributing $T_k$ units over $N_k-1$ ranks, the rank of the $S_k$th unit must lie between $S_k$ and $N_k-1-T_k+S_k$. For every $l\in \{S_k..N_k-1-T_k+S_k\}$, there are $\binom{l-1}{S_k-1}\binom{N_k-1-l}{T_k-S_k}$ ways of distributing those $T_k$ units while having the $S_k$th unit at the $l$th rank. Thus,
\begin{equation}\label{eq:lem_sup5}
\sum\limits_{l=S_k}^{N_k-1-T_k+S_k}\binom{l-1}{S_k-1}\binom{N_k-1-l}{T_k-S_k}=\binom{N_k-1}{T_k}.
\end{equation}
The eighth equality follows from  \eqref{eq:lem_sup4} and \eqref{eq:lem_sup5}. This concludes the proof of \eqref{eq:lem1}.

\medskip
Second, we show that  \eqref{eq:lem_sup0} holds when $\mathcal{P}_k$ is such that $T_k=N_k$. Then, we have
\begin{eqnarray}
E\left(\frac{1}{L_k-1}\sum_{i:Z_{ik}=1}w_{ik}\phi(P_{ik})\middle|~\mathcal{P}_k\right)&=&E\left(\sum_{i=1}^{N_k}\phi(P_{ik})\frac{1}{S_k}1\{R_{ik}\leq S_k\}\middle|~\mathcal{P}_k\right)\nonumber\\
&=&\sum_{i=1}^{N_k}\phi(P_{ik})\frac{1}{S_k}E\left(1\{R_{ik}\leq S_k\}\middle|~\mathcal{P}_k\right)\nonumber\\
&=&\frac{1}{N_k}\sum_{i=1}^{N_k}\phi(P_{ik}).\label{eq:lem2}
\end{eqnarray}
The first equality follows from the definition of $w_{ik}$ and from the fact that if $T_k=N_k$, $L_k=S_k$. The second equality holds because $\phi(P_{ik})$ is a function of $\mathcal{P}_k$, $N_k$ and $S_k$ are non stochastic, and the conditional expectation is linear. The third equality follows from the fact that under Assumption \ref{hyp:random_assignment_of_seats}, if $T_k=N_k$ then conditional on $\mathcal{P}_k$ each applicant has a probability $\frac{S_k}{N_k}$ of having $R_{ik}\leq S_k$. This proves  \eqref{eq:lem2}.  \eqref{eq:lem1} and \eqref{eq:lem2} prove  \eqref{eq:lem_sup0}.

\medskip
We then show that
\begin{equation}\label{eq:lem_sup6}
E\left(\frac{1}{N_k-L_k}\sum_{i:Z_{ik}=0}\phi\left(P_{ik}\right)\middle|~\mathcal{P}_k\right)=\frac{1}{N_k}\sum_{i=1}^{N_k}\phi\left(P_{ik}\right).
\end{equation}
First, we show that  \eqref{eq:lem_sup6} holds when $\mathcal{P}_k$ is such that $T_k<N_k$. Then, we have
\small
\begin{adjustwidth}{-2cm}{-2cm}
\begin{eqnarray}
&&E\left(\frac{1}{N_k-L_k}\sum_{i:Z_{ik}=0}\phi\left(P_{ik}\right)\middle|~\mathcal{P}_k\right)\nonumber\\
&=&\sum_{i=1}^{N_k}\phi(P_{ik})E\left(\frac{1}{N_k-L_k}1\{R_{ik}> L_k\}\middle|~\mathcal{P}_k\right)\nonumber\\
&=&\sum_{i=1}^{N_k}\phi(P_{ik})\sum\limits_{l=S_k}^{N_k-T_k+S_k}\frac{\binom{l-1}{S_k-1}\binom{N_k-l}{T_k-S_k}}{\binom{N_k}{T_k}}\frac{1}{N_k-l}E\left(1\{R_{ik}> l\}\middle|~L_k=l,\mathcal{P}_k\right)\nonumber\\
&=&\sum_{i=1}^{N_k}\phi(P_{ik})\sum\limits_{l=S_k}^{N_k-T_k+S_k}\frac{\binom{l-1}{S_k-1}\binom{N_k-l}{T_k-S_k}}{\binom{N_k}{T_k}}\frac{1}{N_k-l}\left(D_{ik}(1)\frac{T_k-S_k}{T_k}+(1-D_{ik}(1))\frac{N_k-T_k-l+S_k}{N_k-T_k}\right)\nonumber\\
&=&\frac{1}{N_k}\sum_{i=1}^{N_k}\phi(P_{ik})\left(D_{ik}(1)\sum\limits_{l=S_k}^{N_k-T_k+S_k}\frac{\binom{l-1}{S_k-1}\binom{N_k-l}{T_k-S_k}\frac{T_k-S_k}{N_k-l}}{\binom{N_k}{T_k}\frac{T_k}{N_k}}+(1-D_{ik}(1))\sum\limits_{l=S_k}^{N_k-1-T_k+S_k}\frac{\binom{l-1}{S_k-1}\binom{N_k-l}{T_k-S_k}\frac{N_k-T_k-l+S_k}{N_k-l}}{\binom{N_k}{T_k}\frac{N_k-T_k}{N_k}}\right)\nonumber\\
&=&\frac{1}{N_k}\sum_{i=1}^{N_k}\phi(P_{ik})\left(D_{ik}(1)\sum\limits_{l=S_k}^{N_k-T_k+S_k}\frac{\binom{l-1}{S_k-1}\binom{N_k-1-l}{T_k-1-S_k}}{\binom{N_k-1}{T_k-1}}+(1-D_{ik}(1))\sum\limits_{l=S_k}^{N_k-1-T_k+S_k}\frac{\binom{l-1}{S_k-1}\binom{N_k-1-l}{T_k-S_k}}{\binom{N_k-1}{T_k}}\right)\nonumber\\
&=&\frac{1}{N_k}\sum_{i=1}^{N_k}\phi(P_{ik}).\label{eq:lem4}
\end{eqnarray}
\end{adjustwidth}
\normalsize
This derivation follows from arguments similar to those used when deriving  \eqref{eq:lem1}. We only prove the last equality.
Note that 
Assumption \ref{hyp:moretakersthanseats} ensures that $1\leq S_k \leq T_k-1\leq N_k-1$,
thus ensuring that all the quantities that follow are well-defined.
There are $\binom{N_k-1}{T_k-1}$ ways of distributing $T_k-1$ units over $N_k-1$ ranks. The rank of the $S_k$th unit must be included between $S_k$ and $N_k-T_k+S_k$, and for every $l\in \{S_k..N_k-T_k+S_k\}$, there are $\binom{l-1}{S_k-1}\binom{N_k-1-l}{T_k-1-S_k}$ ways of distributing those $T_k-1$ units while having that the $S_k$th unit is at the
 $l$th rank. Therefore,
\begin{equation}\label{eq:lem_sup7}
\sum\limits_{l=S_k}^{N_k-T_k+S_k}\binom{l-1}{S_k-1}\binom{N_k-1-l}{T_k-1-S_k}=\binom{N_k-1}{T_k-1}.
\end{equation}
The last equality in the derivation of  \eqref{eq:lem4} follows from  \eqref{eq:lem_sup5} and \eqref{eq:lem_sup7}.

\medskip
Second, we show that  \eqref{eq:lem_sup6} holds when $\mathcal{P}_k$ is such that $T_k=N_k$. Then, we have
\small
\begin{equation}
E\left(\frac{1}{N_k-L_k}\sum_{i:Z_{ik}=0}\phi(P_{ik})\middle|~\mathcal{P}_k\right)=\sum_{i=1}^{N_k}\phi(P_{ik})\frac{1}{N_k-S_k}E\left(1\{R_{ik}> S_k\}\middle|~\mathcal{P}_k\right)=\frac{1}{N_k}\sum_{i=1}^{N_k}\phi(P_{ik}).\label{eq:lem5}
\end{equation}
\normalsize
This derivation follows from arguments similar to those used when deriving  \eqref{eq:lem2}.  \eqref{eq:lem4} and \eqref{eq:lem5} prove  \eqref{eq:lem_sup6}. QED.

\subsection*{Proof of Lemma \ref{lem:identification}}

We only prove point a), point b) follows from a similar argument.
\begin{eqnarray*}
&&E\left(\frac{1}{K}\sum_{k=1}^K\frac{N_k}{\overline{N}}\left(\frac{1}{L_k-1}\sum_{i:Z_{ik}=1}w_{ik}Y_{ik}-\frac{1}{N_k-L_k}\sum_{i:Z_{ik}=0}Y_{ik}\right)\right)\\
&=&\frac{1}{K}\sum_{k=1}^K\frac{N_k}{\overline{N}}E\left(E\left(\frac{1}{L_k-1}\sum_{i:Z_{ik}=1}w_{ik}Y_{ik}(D_{ik}(1))\middle|~\mathcal{P}_k\right)-E\left(\frac{1}{N_k-L_k}\sum_{i:Z_{ik}=0}Y_{ik}(0)\middle|~\mathcal{P}_k\right)\right)\\
&=&\frac{1}{K}\sum_{k=1}^K\frac{N_k}{\overline{N}}E\left(\frac{1}{N_k}\sum_{i=1}^{N_k}\left[Y_{ik}(D_{ik}(1))-Y_{ik}(0)\right]\right)\\
&=&E\left(\frac{1}{N}\sum_{(i,k)\in \mathcal{I}}\left[Y_{ik}(D_{ik}(1))-Y_{ik}(0)\right]\right).
\end{eqnarray*}
The first equality follows from the linearity of the expectation, from the fact $N_k$ and $\overline{N}$ are not stochastic, from point c) of Assumption \ref{hyp:set-up} and the definitions of $Y_{ik}$ and $D_{ik}$, from the law of iterated expectations, and from the linearity of the conditional expectation.
The second equality follows from Lemma \ref{lem:balancing}, with $\phi(P_{ik})=Y_{ik}(D_{ik}(1))$ for the first conditional expectation, and $\phi(P_{ik})=Y_{ik}(0)$ for the second one.
The third equality follows after some algebra. QED.

\medskip
The proof of Theorem \ref{thm:inference} below makes use of the following lemma, where $O_p(1)$ (resp. $o_p(1)$) stands for a sequence of random variables bounded in probability (resp. converging towards 0 in probability), see, e.g., \cite{van2000}.
\begin{lem}\label{lem_linearization}
Let $(A_K)_{K\in \mathbb{N}}$ and $(B_K)_{K\in \mathbb{N}}$ be two sequences of real numbers such that for every $K$, $B_K\geq C$ for some real number $C>0$, and $\frac{A_K}{B_K}$ converges towards a finite limit. Let $(\widehat{A}_K)_{K\in \mathbb{N}}$ and $(\widehat{B}_K)_{K\in \mathbb{N}}$ be two sequences of random variables such that $\sqrt{K}\left(\widehat{A}_K-A_K\right)=O_p(1)$ and $\sqrt{K}\left(\widehat{B}_K-B_K\right)=O_p(1)$.
Then,
$$\sqrt{K}\left(\frac{\widehat{A}_K}{\widehat{B}_K}-\frac{A_K}{B_K}\right)=\sqrt{K}\frac{1}{B_K}\left((\widehat{A}_K-A_K)-\frac{A_K}{B_K}(\widehat{B}_K-B_K)\right)+o_P(1).$$
\end{lem}
\subsection*{Proof of Lemma \ref{lem_linearization}}

$\sqrt{K}\left(\widehat{A}_K-A_K\right)=O_p(1)$ and $\sqrt{K}\left(\widehat{B}_K-B_K\right)=O_p(1)$ imply that $\widehat{A}_K-A_K=o_p(1)$ and $\widehat{B}_K-B_K=o_p(1)$. Therefore, with probability approaching one, $\max\left(\widehat{A}_K-A_K,\widehat{B}_K-B_K\right)\leq \frac{C}{2}$. Then, Lemma S3 in \cite{deChaisemartin18} implies that with probability approaching one,
\begin{eqnarray*}
&&\left|\sqrt{K}\left(\frac{\widehat{A}_K}{\widehat{B}_K}-\frac{A_K}{B_K}\right)-\sqrt{K}\frac{1}{B_K}\left((\widehat{A}_K-A_K)-\frac{A_K}{B_K}(\widehat{B}_K-B_K)\right)\right|\\
&\leq& \frac{2\left(1+\frac{A_K}{B_K}\right)}{C^2}\max\left(\sqrt{K}(\widehat{A}_K-A_K),\sqrt{K}(\widehat{B}_K-B_K)\right)\max\left(\widehat{A}_K-A_K,\widehat{B}_K-B_K\right).
\end{eqnarray*}
The right hand side of the inequality in the previous display is an $o_p(1)$. With probability approaching one, the left hand side is bounded by an $o_p(1)$, so it is itself an $o_p(1)$. QED.

\subsection*{Proof of Theorem \ref{thm:inference}}

\medskip
\textbf{Proof that $\sqrt{K}\left(\widehat{\Delta}-\Delta_K\right)~{\overset{d}{\longrightarrow}}~ \mathcal{N}\left(0,\sigma^2\right)$}

\medskip
First, notice that
\begin{eqnarray}\label{eq:as_theory0}
\Delta_K&=&
E\left(\frac{T}{E(T)}\frac{1}{T}\sum_{(i,k):D_{ik}(1)=1}\left[Y_{ik}(1)-Y_{ik}(0)\right]\right)\nonumber \\
&=&\frac{E\left(\frac{1}{N}\sum_{(i,k)\in \mathcal{I}}\left[Y_{ik}(D_{ik}(1))-Y_{ik}(0)\right]\right)}{E\left(\frac{1}{N}\sum_{(i,k)\in \mathcal{I}}D_{ik}(1)\right)}\nonumber\\
&=&\frac{E\left(\frac{1}{K}\sum_{k=1}^KRF_k\right)}{E\left(\frac{1}{K}\sum_{k=1}^KFS_k\right)}.
\end{eqnarray}
The first equality follows from point b) of Assumption \ref{hyp:iid}. The second equality follows from some algebra, and from point a) of Assumption \ref{hyp:set-up}. The last equality follows from points a) and b) of Lemma \ref{lem:identification} and from the definitions of $RF_k$ and $FS_k$.

\medskip
Then,
\begin{eqnarray}\label{eq:as_theory1}
&&\sqrt{K}\left(\frac{1}{K}\sum_{k=1}^KRF_k-E\left(\frac{1}{K}\sum_{k=1}^KRF_k\right)\right)\nonumber\\
&=&\frac{\sum_{k=1}^K\left(RF_k-E\left(RF_k\right)\right)}{\sqrt{\sum_{k=1}^KV\left(RF_k\right)}}\sqrt{\frac{1}{K}\sum_{k=1}^KV\left(RF_k\right)}.
\end{eqnarray}
Point a) of Assumption \ref{hyp:set-up} and point a) of Assumption \ref{hyp:iid} ensure that $(RF_k)_{k\in \mathbb{N}}$ is a sequence of independent random variables. Point d) of Assumption \ref{hyp:iid} ensures that for every $k$, the expectation and variance of $RF_k$ exist, and point e) ensures that $(RF_k)_{k\in \mathbb{N}}$ satisfies the Liapunov condition (see, e.g., \citep{billingsley1995}, page 362) for $\delta=2$.
Then, the Liapunov central limit theorem implies that
\begin{eqnarray}\label{eq:as_theory2}
&&\frac{\sum_{k=1}^K\left(RF_k-E\left(RF_k\right)\right)}{\sqrt{\sum_{k=1}^KV\left(RF_k\right)}}~{\overset{d}{\longrightarrow}}~\mathcal{N}(0,1).
\end{eqnarray}
Point e) of Assumption \ref{hyp:iid} implies that
\begin{eqnarray}\label{eq:as_theory3}
&&\underset{K\rightarrow +\infty}{\lim}\sqrt{\frac{1}{K}\sum_{k=1}^KV\left(RF_k\right)}=\sigma_{RF},
\end{eqnarray}
for some real number $\sigma_{RF}$.
Therefore, combining \eqref{eq:as_theory1}, \eqref{eq:as_theory2}, \eqref{eq:as_theory3}, and the Slutsky lemma,
\begin{eqnarray}\label{eq:as_theory4}
&&\sqrt{K}\left(\frac{1}{K}\sum_{k=1}^K\left(RF_k-E\left(\frac{1}{K}\sum_{k=1}^KRF_k\right)\right)\right)~{\overset{d}{\longrightarrow}}~\mathcal{N}(0,\sigma_{RF}^2).
\end{eqnarray}
Similarly, one can show that
\begin{eqnarray}\label{eq:as_theory5}
&&\sqrt{K}\left(\frac{1}{K}\sum_{k=1}^KFS_k-E\left(\frac{1}{K}\sum_{k=1}^KFS_k\right)\right)~{\overset{d}{\longrightarrow}}~\mathcal{N}(0,\sigma_{FS}^2),
\end{eqnarray}
for some real number $\sigma_{FS}$.

\medskip
Finally,
\begin{align*}
&\sqrt{K}\left(\widehat{\Delta}-\Delta_K\right)\\
=&\sqrt{K}\left(\frac{\frac{1}{K}\sum_{k=1}^KRF_k}{\frac{1}{K}\sum_{k=1}^KFS_k}-\frac{E\left(\frac{1}{K}\sum_{k=1}^KRF_k\right)}{E\left(\frac{1}{K}\sum_{k=1}^KFS_k\right)}\right)\\
=&\sqrt{K}\frac{1}{E\left(\frac{1}{K}\sum_{k=1}^KFS_k\right)}\left(\frac{1}{K}\sum_{k=1}^KRF_k-E\left(\frac{1}{K}\sum_{k=1}^KRF_k\right)\right.\\
&\left.-\frac{E\left(\frac{1}{K}\sum_{k=1}^KRF_k\right)}{E\left(\frac{1}{K}\sum_{k=1}^KFS_k\right)}\left(\frac{1}{K}\sum_{k=1}^KFS_k-E\left(\frac{1}{K}\sum_{k=1}^KFS_k\right)\right)\right)+o_P(1)\\
=&\sqrt{K}\frac{1}{E\left(\frac{1}{K}\sum_{k=1}^KFS_k\right)}\left(\frac{1}{K}\sum_{k=1}^KRF_k-E\left(\frac{1}{K}\sum_{k=1}^KRF_k\right)-\Delta\left(\frac{1}{K}\sum_{k=1}^KFS_k-E\left(\frac{1}{K}\sum_{k=1}^KFS_k\right)\right)\right)+o_P(1)\\
=&\frac{FS}{E\left(\frac{1}{K}\sum_{k=1}^KFS_k\right)}\sqrt{K}\left(\frac{1}{K}\sum_{k=1}^K\left(\Lambda_k-E\left(\Lambda_k\right)\right)\right)+o_P(1)~{\overset{d}{\longrightarrow}}~\mathcal{N}(0,\sigma^2).
\end{align*}
The first equality follows from the definitions of $FS_k$ and $RF_k$ and from \eqref{eq:as_theory0}.

\medskip
The second equality follows from the fact $E\left(\frac{1}{K}\sum_{k=1}^KRF_k\right)$, $E\left(\frac{1}{K}\sum_{k=1}^KFS_k\right)$, $\frac{1}{K}\sum_{k=1}^KRF_k$, and $\frac{1}{K}\sum_{k=1}^KFS_k$ satisfy the assumptions of Lemma \ref{lem_linearization}. Indeed, point b) of Lemma \ref{lem:identification}, point c) of Assumption \ref{hyp:iid}, and Assumption \ref{hyp:moretakersthanseats} imply that $E\left(\frac{1}{K}\sum_{k=1}^KFS_k\right)\geq \frac{3}{N^+}>0$. Moreover, point e) of Assumption \ref{hyp:iid} implies that $E\left(\frac{1}{K}\sum_{k=1}^KRF_k\right)/E\left(\frac{1}{K}\sum_{k=1}^KFS_k\right)$ converges towards a finite limit. Finally, it follows from \eqref{eq:as_theory4}, \eqref{eq:as_theory5}, and the fact that convergence in distribution implies boundedness in probability, that
\begin{eqnarray*}
&&\sqrt{K}\left(\frac{1}{K}\sum_{k=1}^KRF_k-E\left(\frac{1}{K}\sum_{k=1}^KRF_k\right)\right)=O_p(1)\\
&&\sqrt{K}\left(\frac{1}{K}\sum_{k=1}^KFS_k-E\left(\frac{1}{K}\sum_{k=1}^KFS_k\right)\right)=O_p(1).
\end{eqnarray*}
\eqref{eq:as_theory0}, point e) of Assumption \ref{hyp:iid}, and \eqref{eq:as_theory5} ensure that $$\frac{1}{E\left(\frac{1}{K}\sum_{k=1}^KFS_k\right)}\left(\Delta-\frac{E\left(\frac{1}{K}\sum_{k=1}^KRF_k\right)}{E\left(\frac{1}{K}\sum_{k=1}^KFS_k\right)}\right)\sqrt{K}\left(\frac{1}{K}\sum_{k=1}^KFS_k-E\left(\frac{1}{K}\sum_{k=1}^KFS_k\right)\right)=o_P(1),$$ hence the third equality. The fourth equality follows from the definition of $\Lambda_k$. The convergence in distribution arrow follows from a reasoning similar to that used to prove \eqref{eq:as_theory4}, and from the Slutsky lemma and the definition of $FS$.

\medskip
\textbf{Proof that $\widehat{\sigma}^2_+~{\overset{p}{\longrightarrow}}~\sigma^2_+\geq \sigma^2.$}

\medskip
By point a) of Assumption \ref{hyp:set-up}, points a), d), and f) of Assumption \ref{hyp:iid}, Kolmogorov's strong law (see, e.g., Theorem 2.3.10 in \citep{sen1993}), and the fact that almost sure convergence implies convergence in probability,
\begin{eqnarray*}
&&\frac{1}{K}\sum_{k=1}^K RF_k-\frac{1}{K}\sum_{k=1}^K E\left(RF_k\right)~{\overset{p}{\longrightarrow}}~0\\
&&\frac{1}{K}\sum_{k=1}^K FS_k-\frac{1}{K}\sum_{k=1}^K E\left(FS_k\right)~{\overset{p}{\longrightarrow}}~0.
\end{eqnarray*}
Then, as under point e) of Assumption \ref{hyp:iid}, $\frac{1}{K}\sum_{k=1}^K E\left(RF_k\right)$ and $\frac{1}{K}\sum_{k=1}^K E\left(FS_k\right)$ converge towards finite limits, the previous display implies that
\begin{eqnarray}\label{eq:as_theory6}
&&\frac{1}{K}\sum_{k=1}^K RF_k~{\overset{p}{\longrightarrow}}~\underset{K\rightarrow +\infty}{\lim}\frac{1}{K}\sum_{k=1}^K E\left(RF_k\right)\nonumber\\
&&\frac{1}{K}\sum_{k=1}^K FS_k~{\overset{p}{\longrightarrow}}~\underset{K\rightarrow +\infty}{\lim}\frac{1}{K}\sum_{k=1}^K E\left(FS_k\right).
\end{eqnarray}
Then, the first point of the theorem and point e) of Assumption \ref{hyp:iid} imply that
\begin{eqnarray}\label{eq:as_theory7}
&&\widehat{\Delta}~{\overset{p}{\longrightarrow}}~\Delta.
\end{eqnarray}
Then, \eqref{eq:as_theory6}, \eqref{eq:as_theory7}, and the continuous mapping theorem imply that
\begin{align}\label{eq:as_theory8}
&\frac{1}{K}\sum_{k=1}^K \widehat{\Lambda}_k=\frac{1}{\frac{1}{K}\sum_{k=1}^K FS_k}\left(\frac{1}{K}\sum_{k=1}^K RF_k-\widehat{\Delta}\frac{1}{K}\sum_{k=1}^K FS_k\right)~{\overset{p}{\longrightarrow}}~\underset{K\rightarrow +\infty}{\lim}\frac{1}{K}\sum_{k=1}^K E\left(\Lambda_k\right).
\end{align}
Similarly, one can show that
\begin{align}\label{eq:as_theory9}
&&\frac{1}{K}\sum_{k=1}^K \widehat{\Lambda}_k^2~{\overset{p}{\longrightarrow}}~\underset{K\rightarrow +\infty}{\lim}\frac{1}{K}\sum_{k=1}^K E\left(\Lambda_k^2\right).
\end{align}
Then, \eqref{eq:as_theory8}, \eqref{eq:as_theory9}, and the continuous mapping theorem imply that
\begin{equation}\label{eq:as_theory10}
\widehat{\sigma}^2_+=\frac{1}{K}\sum_{k=1}^K \widehat{\Lambda}_k^2-\left(\frac{1}{K}\sum_{k=1}^K \widehat{\Lambda}_k\right)^2~{\overset{p}{\longrightarrow}}~\sigma^2_+.
\end{equation}
Finally, the convexity of $x\mapsto x^2$ implies that $\frac{1}{K}\sum_{k=1}^K E\left(\Lambda_k\right)^2\geq \left(\frac{1}{K}\sum_{k=1}^K E\left(\Lambda_k\right)\right)^2,$
so $\sigma^2_+\geq \sigma^2$. QED.

\medskip
Theorem \ref{thm:inconsistencyEO} relies on Assumption \ref{hyp:plim_EO} below. Let
\begin{align*}
&RF^E_k=N_k\frac{L_k}{N_k}\left(1-\frac{L_k}{N_k}\right)\left(\frac{1}{L_k}\sum_{i:Z_{ik}=1}Y_{ik}-\frac{1}{N_k-L_k}\sum_{i:Z_{ik}=0}Y_{ik}\right),\\
&FS^E_k=N_k\frac{L_k}{N_k}\left(1-\frac{L_k}{N_k}\right)\frac{1}{L_k}\sum_{i:Z_{ik}=1}D_{ik}.
\end{align*}
\begin{hyp}\label{hyp:plim_EO}
(Technical assumptions to derive the probability limit of $\widehat{\beta}^{E}_{FE}$)

\medskip
For every $k$ $E\left(\left(RF^E_k\right)^2\right)<+\infty$; 
    $\sum_{k=1}^{+\infty}\frac{V\left(RF^E_k\right)}{k^2}<+\infty$; $\frac{1}{K}\sum_{k=1}^KE\left(\frac{S_k\left(N_k-S_k\frac{N_k+1}{T_k+1}\right)}{N_k}\right)$, $\frac{1}{K}\sum_{k=1}^KE\left(\frac{S_k\left(N_k-T_k\frac{N_k+1}{T_k+1}\right)}{N_k}\left[\frac{1}{T_k}\sum_{i:D_{ik}(1)=1}Y_{ik}(0)-\frac{1}{N_k-T_k}\sum_{i:D_{ik}(1)=0}Y_{ik}(0)\right]\right)$, and \\
    $\frac{1}{K}\sum_{k=1}^K E\left(\frac{S_k\left(N_k-S_k\frac{N_k+1}{T_k+1}\right)}{N_k}\frac{1}{T_k}\sum_{i:D_{ik}(1)=1}\left[Y_{ik}(1)-Y_{ik}(0)\right]\right)$ have finite limits when $K\rightarrow +\infty$.
\end{hyp}

\subsection*{Proof of Theorem \ref{thm:inconsistencyEO}}

First,
\begin{align}\label{eq:E(L)}
E(L_k|\mathcal{P}_k)=&\sum_{l=S_k}^{N_k-T_k+S_k}l\frac{\binom{l-1}{S_k-1}\binom{N_k-l}{T_k-S_k}}{\binom{N_k}{T_k}}\nonumber\\
=&S_k\frac{N_k+1}{T_k+1}\sum_{l=S_k}^{N_k-T_k+S_k}\frac{\binom{l}{S_k}\binom{N_k+1-(l+1)}{T_k+1-(S_k+1)}}{\binom{N_k+1}{T_k+1}}\nonumber\\
=&S_k\frac{N_k+1}{T_k+1}.
\end{align}
This derivation follows from arguments similar to those used when deriving  \eqref{eq:lem1}.

\medskip
Then, it follows from the fact that a 2SLS coefficient with one endogenous variable and one instrument is equal to the ratio of the reduced form and first stage coefficients, from Equation (3.3.7) in \cite{angrist2008}, and from the definitions of $RF^E_k$ and $FS^E_k$, that
\begin{equation}\label{eq:beta_EO}
\widehat{\beta}^{E}_{FE}=\frac{\frac{1}{K}\sum_{k=1}^KRF^E_k}{\frac{1}{K}\sum_{k=1}^KFS^E_k}.
\end{equation}
For every $k$,
\begin{align}\label{eq:as_theory11}
&E\left(RF^E_k\right)\nonumber\\
=&E\left(\left(1-\frac{L_k}{N_k}\right)\sum_{i=1}^{N_k} Y_{ik}(D_{ik}(1))1\{R_{ik}\leq L_k\}-\frac{L_k}{N_k}\sum_{i=1}^{N_k}Y_{ik}(0)(1-1\{R_{ik}\leq L_k\})\right)\nonumber\\
=&E\left(\left(1-\frac{L_k}{N_k}\right)\sum_{i=1}^{N_k} Y_{ik}(D_{ik}(1))E(1\{R_{ik}\leq L_k\}|L_k,\mathcal{P}_k)-\frac{L_k}{N_k}\sum_{i=1}^{N_k}Y_{ik}(0)(1-E(1\{R_{ik}\leq L_k\}|L_k,\mathcal{P}_k))\right)\nonumber\\
=&E\left(\left(1-\frac{L_k}{N_k}\right)\sum_{i=1}^{N_k} Y_{ik}(D_{ik}(1))\left(D_{ik}(1)\frac{S_k}{T_k}+(1-D_{ik}(1))\frac{L_k-S_k}{N_k-T_k}\right)\right.\nonumber\\
&\left.-\frac{L_k}{N_k}\sum_{i=1}^{N_k}Y_{ik}(0)\left(D_{ik}(1)\frac{T_k-S_k}{T_k}+(1-D_{ik}(1))\frac{N_k-T_k-L_k+S_k}{N_k-T_k}\right)\right)\nonumber\\
=&E\left(\frac{(N_k-L_k)S_k}{N_k}\frac{1}{T_k}\sum_{i:D_{ik}(1)=1}Y_{ik}(1)-\frac{L_k(T_k-S_k)}{N_k}\frac{1}{T_k}\sum_{i:D_{ik}(1)=1}Y_{ik}(0)\right.\nonumber\\
&\left.+\frac{(N_k-L_k)(L_k-S_k)-L_k(N_k-T_k-L_k+S_k)}{N_k}\frac{1}{N_k-T_k}\sum_{i:D_{ik}(1)=0}Y_{ik}(0)\right)\nonumber\\
=&E\left(\frac{(N_k-L_k)S_k}{N_k}\frac{1}{T_k}\sum_{i:D_{ik}(1)=1}\left[Y_{ik}(1)-Y_{ik}(0)\right]\right.\nonumber\\
&\left.+\frac{N_kS_k-L_kT_k}{N_k}\left(\frac{1}{T_k}\sum_{i:D_{ik}(1)=1}Y_{ik}(0)-\frac{1}{N_k-T_k}\sum_{i:D_{ik}(1)=0}Y_{ik}(0)\right)\right)\nonumber\\
=&E\left(\frac{S_k\left(N_k-S_k\frac{N_k+1}{T_k+1}\right)}{N_k}\frac{1}{T_k}\sum_{i:D_{ik}(1)=1}\left[Y_{ik}(1)-Y_{ik}(0)\right]\right.\nonumber\\
&\left.+\frac{S_k\left(N_k-T_k\frac{N_k+1}{T_k+1}\right)}{N_k}\left(\frac{1}{T_k}\sum_{i:D_{ik}(1)=1}Y_{ik}(0)-\frac{1}{N_k-T_k}\sum_{i:D_{ik}(1)=0}Y_{ik}(0)\right)\right).
\end{align}
The first equality follows from the definition of $RF^E_k$ and some algebra. The second equality follows from the law of iterated expectations and the linearity of the conditional expectation. The third equality follows from \eqref{eq:lem_sup2} and \eqref{eq:lem_sup3}. The fourth and fifth equality follow from some algebra. The last equality follows from the law of iterated expectations, the linearity of the conditional expectation, and \eqref{eq:E(L)}.

\medskip
Similarly, one can show that for every $k$,
\begin{align}\label{eq:as_theory12}
&E\left(FS^E_k\right)=E\left(\frac{S_k\left(N_k-S_k\frac{N_k+1}{T_k+1}\right)}{N_k}\right).
\end{align}

\medskip
Equations \eqref{eq:as_theory11} and \eqref{eq:as_theory12} combined with Assumption \ref{hyp:plim_EO} imply that $\frac{1}{K}\sum_{k=1}^KE\left(RF^E_k\right)$ and $\frac{1}{K}\sum_{k=1}^KE\left(FS^E_k\right)$ converge towards finite limits when $K\rightarrow +\infty$.
Then, one can use a reasoning similar to that used to prove \eqref{eq:as_theory6} to show that
\begin{equation}\label{eq:as_theory13}
\widehat{\beta}^{E}_{FE}~{\overset{p}{\longrightarrow}}~\frac{\underset{K\rightarrow +\infty}{\lim}\frac{1}{K}\sum_{k=1}^K E\left(RF^E_k\right)}{\underset{K\rightarrow +\infty}{\lim}\frac{1}{K}\sum_{k=1}^K E\left(FS^E_k\right)}.
\end{equation}
The result follows from plugging \eqref{eq:as_theory11} and \eqref{eq:as_theory12} into \eqref{eq:as_theory13}. QED.

\end{document}